\documentclass{aa}
\usepackage{graphics,epsfig}

\def\simless{\mathbin{\lower 3pt\hbox
     {$\rlap{\raise 5pt\hbox{$\char'074$}}\mathchar"7218$}}}   
\def\simmore{\mathbin{\lower 3pt\hbox
     {$\rlap{\raise 5pt\hbox{$\char'076$}}\mathchar"7218$}}}   

\newcommand{\be}{\begin{equation}}
\newcommand{\ee}{\end{equation}}

\begin{document}

\title{Spectra of black-hole binaries
in the low/hard state: from radio to X-rays}

\subtitle{}

\author
{Dimitrios Giannios\inst{1,2}}

\institute{
University of Crete, Physics Department, P.O. Box 2208, 710 03,
Heraklion, Crete, Greece
\and Foundation for Research and Technology-Hellas, 711 10, Heraklion,
Crete, Greece}

\authorrunning{Giannios}
\titlerunning{Spectra and variability of black-hole binaries
in the low state}

\offprints{giannios@mpa-garching.mpg.de}

\date{Received: \\
Accepted: \\}

\abstract{

We propose a jet model for the low/hard state of Galactic black-hole X-ray sources that
can explain the energy spectra from radio to X-rays.  The 
model assumes that i) there is a magnetic field along the
axis of the jet; ii) the electron density in the jet drops inversely proportional to 
distance; and iii) the electrons in the jet follow a power law distribution function.  We have 
performed Monte Carlo simulations of Compton upscattering of soft photons from the 
accretion disk and have found power-law high-energy spectra with photon-number index
in the range 1.5 - 2 and  cutof\mbox{}f at a few hundred keV. The spectrum at long wavelengths 
(radio, infrared, optical) is modeled to come 
from synchrotron radiation of the energetic electrons in the jet. We find flat to inverted 
radio spectra that extend from the radio up to about the optical band. For magnetic
field strengths of the order of $10^5-10^6$G at the base of the jet, the calculated
spectra agree well in slope and flux with the observations. Our
model has the advantage over other existing models that it also explains  many of the
existing timing data such as the time lag spectra, the hardening of the
power density spectra  and the narrowing of the autocorrelation
function with increasing photon energy.

}

\maketitle

\section{Introduction} \label{intro}

Galactic black-hole binaries are classified by their X-ray features such as the
strength and temperature of the soft multi color black-body component (Mitsuda et al.
1984), the 
hard X-ray emission, the X-ray luminosity and the timing properties.
Several states have been identified to characterize black-hole accretion
(for reviews see Tanaka \& Lewin 1995; van der Klis 1995; Nowak 1995; Poutanen 1998; 
 McClintock \& Remillard 2005).
A source is in the so-called low/hard state when it exhibits low X-ray luminosity 
(less than a few \% the
Eddington luminosity $L_{\rm E}$), weak or absent thermal component, dominant
 hard X-ray emission with a cutof\mbox{}f at $\sim$ a few hundred keV
and high rms variability.

The soft thermal component is widely accepted to come from 
an optically thick, geometrically thin accretion disk (Shakura \& Sunyaev 1973), 
while the hard X-ray tail is usually assumed to come from inverse Compton 
scattering of soft photons off energetic electrons (e.g. Sunyaev \& Titarchuk 1980).
The location of the energetic electrons and, more generally, the accretion
geometry are poorly understood. One possibility is that the disk extends
all the way down to the last stable circular orbit and that there is a hot rarefied
``corona'' that lies above and below the thin disk (Galeev et al. 1979; Haardt \& Maraschi 
1993; Haardt et al. 1994; Stern et al. 1995; Poutanen \& Fabian 1999).
It is also possible that the disk is truncated at a larger radius and 
the inner region is filled with optically thin, geometrically thick,
two-temperature plasma (Shapiro et al. 1976; Ichimaru 1977; Rees et al. 1982;
Narayan \& Yi 1994; Abramowicz et al. 1995; Esin et al. 1998).

During recent years, evidence has mounted that, when in the low/hard state,
X-ray binaries exhibit steady core radio emission (Hjellming \& Han 1995; 
Mirabel et al. 1998; Fender 2001; Gallo et al. 2003).  Moreover, in some cases 
 a jet-like structure has been resolved (Mirabel et al. 1992; Dhawan et al. 2000; 
Stirling et al. 2001; Mart\'{\i}  et al. 2002; Fuchs et al. 2003). The spectrum
in the radio is flat to inverted and seems to extend to 
 the infrared or even beyond (see, e.g., Hannikainen et al. 1998; 
Fender et al. 2000; Corbel et al. 2000). Because of the high brightness temperatures,  
non-thermal spectra and, in some cases, high degree of polarization, radio
emission is believed to come from synchrotron radiation of relativistic 
electrons in a jet. Furthermore, an interesting radio/X-ray correlation
has been established to hold through simultaneous radio and X-ray observations
(Gallo et al. 2003)  which can be extended to include the mass of the black
hole and to define the ``fundamental plane'' of black hole activity (Merloni et al. 2003; 
see also Falcke et al. 2004).

Inverse Compton scattering by 
relativistic electrons in a jet has been proposed as a mechanism for the 
production of X-rays and $\gamma$-rays in 
X-ray binaries (Band \& Grindlay 1986; Levinson \& Blandford 1996; Georganopoulos et
al. 2002; Romero et al. 2002). The possibility that optically thin synchrotron
emission from the jet results in the hard X-ray tail has
also been explored (Markoff et al. 2001;  Vadawale et al. 2001; Corbel \& Fender 2002; 
Markoff et al. 2003;  see also Markoff \& Nowak 2004 for constraints on
 jet models via reflection).

Recently, Reig et al. (2003) (hereafter Paper I) proposed a jet model
that can explain the X-ray energy spectra and the dependence of time lags
on Fourier frequency (see for example Miyamoto et al. 1988; Nowak et al. 1999; 
Ford et al. 1999) in terms of inverse Compton scattering of soft,
disk photons by energetic electrons in the jet.   
The density of the electrons in this simple jet model is assumed to drop 
inversely proportional to the vertical distance $z$ from the black hole. Giannios et
al. (2004) (hereafter Paper II; see also Kylafis et al. 2004) further assumed that the 
electrons close to the core of the jet
are more energetic than those at its periphery and showed that both the
hardening of the high-frequency power spectra and the narrowing of the 
auto-correlation  function with photon energy observed in Cygnus X-1 (Nowak et al. 1999;
Revnivtsev et al. 2000; Maccarone et al. 2000) are reproduced.

Although successful in explaining a large number of spectral and timing properties
in the X-ray domain, Papers I, II did not deal with the part of the electromagnetic 
spectrum that revealed the existence of the jet in the first place, i.e. the radio.
Radio observations on the other hand contain valuable information (spectral slope, flux)
that can place constrains on any jet model.  
In this work, we model the radio emission
in terms of synchrotron radiation of the electrons in the jet. The main modification
with respect to the jet model used in Papers I, II is that we take the electrons
to have a power law energy distribution function. We first make sure that this
modification does not alter any of the previously derived results in the X-ray domain.
This puts significant constraints on the index of the power law energy distribution
of the electrons. Then we show that flat to inverted radio spectra are naturally reproduced 
by the model. Finally, we apply the model to the broad band spectra of XTE J1118+480 and
Cygnus X-1.

\section{The model}
\subsection{Description of the jet} 

The jet model that we will use here is built on the model of Paper I.
Here we describe the characteristics of the jet.
We assume that the jet is accelerated close to its launching region and that
it has constant velocity $v_{\parallel}$ (say along the $z$-axis). Furthermore,
we assume that the electron density in the jet drops inversely proportional with 
distance $z$
\be
n(z)=n_0\frac{z_0}{z},
\label{density}
\ee
where $n_0$ is the density and $z_0$ is the height at the base of the jet respectively. 
Mass conservation $\dot M\propto v_{\parallel}n(z)R^2(z)$ then determines the polar radius
$R$ of the outer edge of the jet as a function of $z$
\be
R(z)=R_0\big(\frac{z}{z_0}\big)^{1/2},
\label{Rz}
\ee
where $R_0$ is the radius at the base of the jet.
The Thomson optical depth along the axis of the jet is given by the integration of
$d\tau=n(z)\sigma_{\rm T} dz$ from $z_0$ to $H$ (where $H$ is the extent of the jet)
\be
\tau_{\parallel}=n_0\sigma_Tz_0\ln \big(\frac{H}{z_0}\big).
\label{taupar}
\ee
 
In the previous works (Papers I, II) the magnetic field had been assumed, for simplicity,
to be parallel 
to the axis of the jet ($z$-axis) and homogeneous. Here, we will keep $\vec B\parallel z$ 
but take its strength to vary with $z$ as dictated by magnetic flux conservation along 
the jet: $B(z)\pi R^2(z)=$ const. So we have for the $z$-dependence of the magnetic field
\be 
B(z)=B_0\frac{z_0}{z}.
\label{bfield}
\ee

The main modification with respect to Papers I and II is in the  
energy distribution of the electrons. In Paper I, the electrons had been taken monoenergetic 
with their velocities to have a constant spiraling perpendicular component $v_{\perp}$. 
In Paper II, $v_{\perp}$ had been assumed to drop linearly with polar distance by a factor 
of a few (resulting in a jet with a core ``hotter''
than its periphery), explaining the hardening of the high-frequency power spectrum with
increasing photon energy in Cygnus X-1 (Nowak et al. 1999).

Here we will assume that the electrons have a distribution of $v_{\perp}$ that extends 
to ultra-relativistic electron velocities. We will describe the distribution
in terms of the Lorentz factor 
\be
\gamma=1/\sqrt{1-(v_{\parallel}^2+v_{\perp}^2)/c^2}
\label{gamma}
\ee  
and will assume -as it is customary in many jet models (e.g. Blandford \& K\"onigl
1979)- a power law form
\be
N(\gamma)d\gamma \propto {\gamma}^{-\alpha}d\gamma. 
\ee
The distribution extends from
$\gamma_{\rm min}$ to $\gamma_{\rm max}$ with $\gamma_{\rm max}\gg \gamma_{\rm min}$.
In total, three parameters are needed to determine the electron distribution
inside the jet i.e.  the index
$\alpha$, $\gamma_{\rm min}$, and $\gamma_{\rm max}$.
Since the distribution itself is expressed in terms
of the Lorentz factor $\gamma$, we prefer to keep $\gamma_{\rm min}$ as a parameter  
 instead of $v_{\perp}$ (of course these quantities are related through Eq. (\ref{gamma})).

With this model at hand, the task is to explore its spectral properties in 
essentially the whole electromagnetic spectrum so as to compare with those observed in
black-hole candidates in the low/hard state. Synchrotron emission from the energetic 
electrons inside the magnetic field given by Eq. (\ref{bfield}) 
will be shown to 
dominate from the radio up to about the optical wavelengths, while we model
the hard X-ray part of the spectrum as a result of inverse Compton scattering
of a soft-photon input at the base of the jet by the same population of electrons.    
We have used the Monte Carlo method  (Pozdnyakov et al. 1983) to simulate inverse 
Compton scattering.

\subsection{Simulations of inverse Compton scattering in the jet} \label{code}

Our code is very similar to the one described in Papers I and II. Here, 
we describe only the emission of the soft photons.  Multi-color black-body
photons (i.e. accretion disk photons) of maximum temperature 
$T_{\rm bb}$ are assumed to be emitted from the underlying disk with an
upward isotropic distribution and temperature that scales with polar
radius as $T\propto \rho^{-3/4}$ (Shakura \& Sunyaev 1973). Furthermore, 
synchrotron flux from the jet also contributes to the soft photon input
to be Comptonized and is taken into account when the model is applied to observations
(see Sect. 6).

\section{Model parameters}

The present model has several free parameters, most of which refer to physical
quantities of the jet. These are: 
the temperature $T_{bb}$, the extent $H$ of the jet,
the radius $R_0$ and the height $z_0$ of the base of the jet, 
the Thomson optical depth ${\tau}_{\parallel}$ along the axis of 
the jet, the velocity of the jet $v_{\parallel}$ and the strength
of the magnetic field at the base of the jet $B_0$. The electron
distribution is specified through the minimum Lorentz factor
 $\gamma_{\rm min}$,  the maximum Lorentz factor 
$\gamma_{\rm max}$ and the exponent of the electron distribution $\alpha$ (see Eqs. (5) and
(6)). 

The values of the parameters that reproduce quite well several properties of black-hole
candidates in the low/hard state (with special emphasis to Cygnus X-1) are called 
{\it reference values} in this work and they are:
$kT_{\rm bb} = 0.2$ keV, $H=5\cdot10^8 r_g$, 
$R_0=100 r_g$, $z_0=5r_g$, $\tau_{\parallel}=2.5$, $v_{\parallel}=0.8 c$, 
$B_0=3\cdot 10^5$ G, $\gamma_{\rm min}=2.1$, 
$\gamma_{\rm max}=500$ and $\alpha=4$, 
where $r_g = GM/c^2 \simeq  1.5 \times 10^6$ cm corresponds to  
the gravitational radius of a 10 solar-mass black hole.    

Since the jet is mildly relativistic, the results also depend on the angle $\theta$
of observation with respect to the jet axis. As in Paper II, we will focus in an
intermediate range of observing angles $0.2<\cos \theta<0.6$. Practically, for
the Monte Carlo simulation this means that we count only photons that leave the
jet in this range of angles.    

Despite the fact that the parameter space is rather large, all the parameters
can be constrained since the model is tested against (and succeeds in reproducing)
a large number of timing and spectral properties of black-hole X-ray binaries.
It is encouraging  that none of the parameters needs to be fine tuned 
in order to match the observations. The last statement will be quantified 
in the next sections, where the results are presented.

\subsection{The ejection rates}

Using the reference values, one can calculate the number density $n_0$ of the
electrons (plus possibly positrons) at the base of the jet (Eq. (3)). Assuming a neutral jet
and $f$ electrons per proton (which also means $f-1$ positrons), the ejection rate in the 
jet is given by the expression
\be
\dot M= n_0 \pi R_0^2 v_{\parallel} (m_{\rm e}+\frac{m_{\rm p}}{2f-1}).
\label{mdot}
\ee 

In the case of absence of pairs (i.e. $f=1$) and  using the reference values
of the parameters one finds an ejection rate
$\dot M=8.2\times 10^{19} \quad \rm g\cdot sec^{-1}$. This value is super Eddington but
not necessarily unrealistic. 
In the case, however, of a pair
dominated jet (i.e. $f\gg 1$) the ejection rate can be lower by more than three
 orders of magnitude and much more ``reasonable''.  At this point, 
our knowledge on the composition of the jet is poor and both baryon and pair dominated 
jets are allowed by observations (see for example Fender 2003).

\subsection{The spectral index of the electron distribution}

 In this work we treat the power-law index of the electron distribution $\alpha$ as 
a free parameter and we do not address the issue of particle acceleration
in the jet.  However, the values of $\alpha\simmore 3$ (see next Section) that we use in 
this work are unusually large (with respect to  what is expected from shock 
acceleration for example). Here, we show  that $\alpha$ is {\emph not} the  
spectral index of the injected electrons but that of the (steady-state) emitting electrons
after taking into account radiative (synchrotron and inverse Compton) cooling.
 
Defining the ``cooling Lorentz factor'' $\gamma_{\rm c}$ of the electrons as the Lorentz 
factor at which the radiative cooling time scale $t_{\rm rad}$ equals the dynamical time scale 
$t_{\rm dyn}$, where 
\be
t_{\rm dyn}=\frac{z}{c}=t_{\rm rad}=\frac{t_{\rm syn}t_{\rm Compton}}{t_{\rm syn}+t_{\rm 
Compton}},
\ee
one can estimate  $\gamma_{\rm c}$. The synchrotron cooling time $t_{\rm syn}$ can be 
calculated straightforwardly with the use of Eq. (4), while the Compton cooling time 
can be calculated numerically with the Monte Carlo method (see Sect. 2.2).  
For the reference 
values of the parameters of the jet it is estimated to be $\gamma_{\rm c}\simeq$ 
a few $\simeq \gamma_{\rm min}$. So, it is approximately correct to 
assume that the whole electron distribution is in the fast cooling regime and can be
modeled as a power-law.

\section{Results in the X-ray domain}

In this section, we show our results concerning the spectral properties 
of the electromagnetic radiation in the X-rays. The temporal properties are 
discussed very briefly since a detailed description has already been given in
Paper II.

\begin{figure}
\resizebox{\hsize}{!}{\includegraphics[angle=270]{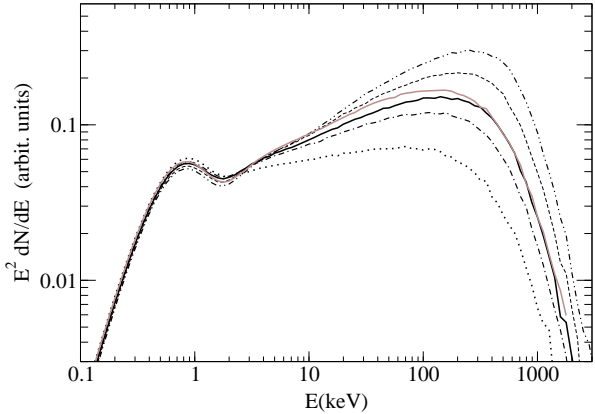}}
\caption[]{Emergent photon-number spectrum multiplied by $E^2$ 
from our jet model for different values of 
the minimum Lorentz factor $\gamma_{\rm min}$ of the electron distribution.
The solid curve corresponds to the reference values of the
parameters. The dotted,  dash-dotted,  dashed and  dash-double-dotted curves
correspond to
$\gamma_{\rm min}=1.8$, $2$, $2.3$, and
$2.5$ respectively. The rest of the parameters are kept at their
reference values. The solid gray curve corresponds to the illustrative example
 where $\gamma_{\rm min}$ varies with $z$ according to Eq. (\ref{gammaz}).} 

\label{spectrum2}
\end{figure}
\begin{figure}
\resizebox{\hsize}{!}{\includegraphics[angle=270]{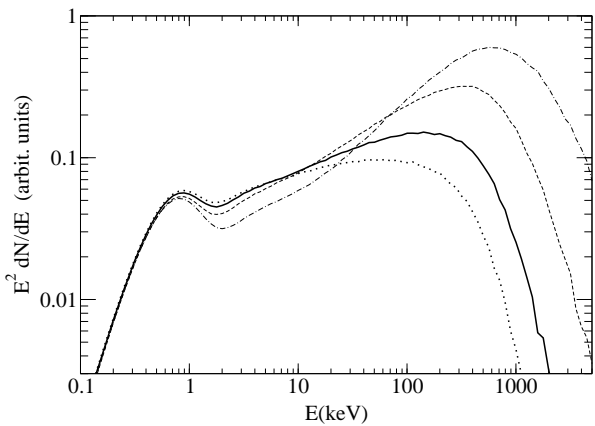}}
\caption[]{Emergent photon-number spectrum multiplied by $E^2$ 
from our jet model for different values of the power-law index $\alpha$.
The dotted, solid, and dashed curves correspond
to $\alpha=5$,  $4$, $3$ respectively. 
 We also plot (dash-dotted curve) the case where $\alpha=2$ and
$\gamma_{\rm min}=1.67$. The rest of the parameters are kept to their reference
values.}

\label{spectrum3}
\end{figure}

\subsection{Energy spectra} \label{espec}
   
In Fig. 1 (solid curve) we plot the energy spectrum of the emerging 
radiation as calculated by our Monte Carlo simulation for the reference 
values of the parameters (see Sect. 3)  and for a black-body soft photon input
\footnote {Since in this Section we explore how the model parameters
affect the X-ray spectra (and mainly the X-ray slope and high energy 
cutof\mbox{}f), the energy distribution of the soft photon input is not very
important. In Sect. 6, however, where the model is applied to data, all
relevant soft photon sources such as multi-temperature disk and synchrotron
photons are taken into account.}.    
 Two components are evident in this plot. Below $\sim$ 1 keV
the dominant component is the soft photon input, while for $E\simmore 1$ keV 
the Comptonized component dominates.
The latter component can be well fitted with a power law of photon-number
index $ \Gamma\simeq 1.7$ with an exponential
cutof\mbox{}f at $E_{\rm cut}\simeq 300$ keV. 

In Papers I, II we have explored how the spectrum depends on some of the parameters
such as the optical depth along the jet $\tau_{\parallel}$ or the extent of the
base of the jet $R_0$. The result was that increasing $\tau_{\parallel}$ or $R_0$,
while keeping the rest of the parameters fixed, the spectrum becomes harder but 
with essentially the same energy cutof\mbox{}f. This is expected since in both cases
there is an increase of the mean number of scatterings that the electrons experience
during their random walk inside the jet and gain more energy on average. The cutof\mbox{}f, 
on the other hand, is mainly determined by the energetics of the electrons which does not
depend on $\tau_{\parallel}$ or $R_0$. The same conclusions have been checked to 
hold in this version of the model. 

Of more interest is to explore how the energetics of the electrons influence
the emerging spectrum. For a power-law distribution of electrons, both
$\gamma_{\rm min}$ and the index $\alpha$ are relevant quantities for this 
exploration. For $\alpha$ large enough, as is the case here (see below), 
the exact value of $\gamma_{\rm max}$ is not important.
It is qualitatively expected that by increasing $\gamma_{\rm min}$ or making the
distribution flatter (i.e. decreasing $\alpha$), the spectrum becomes harder
and the exponential cutoff appears at higher energies. 

In Fig. \ref{spectrum2} we plot the emerging spectrum for different 
values of $\gamma_{\rm min}$, while the rest of the parameters are kept at their 
reference values. One can clearly see the hardening of the spectrum with increasing
$\gamma_{\rm min}$.  More quantitatively, by increasing  $\gamma_{\rm min}$ in the
range $1.8-2.5$, the photon number index $\Gamma$ of the hard X-ray slope hardens from 
$\sim 1.9$ to $\sim 1.5$ which is well within the observed range for black-hole binaries 
in the low/hard state.  For simplicity, we have assumed that 
$\gamma_{\rm min}$ is constant along the jet. This, however,  may not be the case 
and $\gamma_{\rm min}$ may vary as a function of $z$. We have checked that if $\gamma_{\rm min}$
changes rather slowly with $z$, the resulting spectra do not change much.
As an example, in Fig. \ref{spectrum2} we plot the X-ray spectrum
for
\be
\gamma_{\rm min}=\gamma_1(1+0.8\sqrt{z_0/z}),
\label{gammaz}
\ee
where $\gamma_1=1/\sqrt{1-v_{\parallel}^2/c^2}$.     

We also show (in Fig. \ref{spectrum3}) how the emergent spectrum depends on
the index $\alpha$ of the electron distribution. As $\alpha$ decreases, the spectrum
gets harder with a cutof\mbox{}f at higher energies. Notice that for $\alpha=2$ we
use the minimum value one could possibly have for $\gamma_{\rm min}$, namely
$\gamma_{\rm min}=1/\sqrt{1-v_{\parallel}^2/c^2}=1.67$ (see Eq. (5)). 
One important conclusion that can be drawn from
this Figure is that if the electron distribution is rather flat (say $\alpha\sim 2$),
the high energy cutoff becomes too high ($E_{\rm cut}\simeq 1$ MeV or more)
in comparison with the one typically observed in black-hole binaries in the low/hard
state ($\sim$ a few hundred keV) for any choice of $\gamma_{\rm min}$. 
This consideration requires  $\alpha\simmore 3$ if this model is to agree with spectral 
observations. As we have already discussed in Sect. 3.2., such a steep spectral
index is the result of fast radiative (synchrotron and inverse Compton) cooling 
of the electrons.

\subsection{Temporal domain}   

In Papers I and II, it has been shown that the jet model can account for a number of 
temporal properties of X-ray black-hole binaries. Since in this work we have modified the 
energetics of the electrons, it is important to check whether the previously derived
results are altered or not. For example the Fourier frequency dependence of the time lags
between two different energy bands is found to remain essentially the same as that shown 
in Papers I and II, in accordance with observations (see for example Nowak et a. 1999; 
Ford et al. 1999). 
Furthermore, the width of the autocorrelation function has been checked to decrease with photon
energy (as pointed out observationally by Maccarone et al. 2000). Finally, it was shown in Paper 
II that, assuming that more energetic electrons lie in the core of the jet in comparison to 
its periphery, a hardening of the high-frequency power spectrum with increasing photon
energy is expected. If we assume -in the context of the current model- that
the minimum Lorentz factor  $\gamma_{\rm min}$ decreases with polar distance, a similar
hardening of the power spectrum is derived.

\section{Radio emission from the jet}

The existence of a steady jet when a black-hole binary is in the low/hard 
state has been strongly suggested through radio observations (Hjelling \& Han 1995;
Mirabel \& Rodr\'\i guez 1999; Fender 2001). Furthermore, in 
some occasions a jet like structure has even been resolved  (Mirabel et al. 1992; 
Dhawan et al. 2000; Stirling et al. 2001; Mart\'{\i}  et al. 2002; Fuchs et al. 2003). 
Our simple jet model has succeeded in explaining a large number of spectral and 
timing properties of these sources in the X-ray region in terms of inverse Compton scattering
of soft photons by energetic electrons in the jet. Here we will focus on
longer wavelengths and explore the predictions of our model in this region
of the electromagnetic spectrum.

\subsection{The spectral slope in the radio}

The coexistence in the jet of energetic electrons and 
magnetic fields indicates synchrotron emission as the dominant 
radiative mechanism in the radio. The study of radio emission is
more straightforward in a frame that is comoving with the emitting
medium (i.e. the jet). Suppose that an electron has a Lorentz factor
$\gamma$ in the lab frame (i.e. the rest frame of the black hole). Then a Lorentz 
transformation gives the Lorentz factor in the comoving frame     
\be
\gamma_{\rm co}=\sqrt{1-v_{\parallel}^2/c^2}\gamma.      
\ee
Since we have assumed that $v_{\parallel}$ is constant in the jet,
$\gamma_{\rm co}\propto \gamma$. Stated in other words, we still 
have a power law distribution for the electrons in the comoving frame, namely
\be
N(\gamma_{\rm co})d\gamma_{\rm co}=C\gamma_{\rm co}^{-\alpha}
d\gamma_{\rm co}.
\label{distr}
\ee
{\it For the rest of this work, and for simplicity in the notation,
 we will drop the subscript} ``co'' {\it from the Lorentz factor.}
The constant $C$ can be found if we integrate the last expression
from $\gamma_{\rm min}$ to $\gamma_{\rm max}\gg \gamma_{\rm min}$ and 
equate it to the comoving electron density. Doing so we have
\be
C=n\sqrt{1-v_{\parallel}^2/c^2}(\alpha-1)\gamma_{\rm min}^{\alpha-1},
\ee
where we have also assumed that $\alpha>1$, which certainly holds, and
$n$ is given by Eq. (1).

The properties of the jet (i.e. density, magnetic field
strength, energetics of the electrons) may vary only along the axis of the jet 
but not along the perpendicular direction. As is customary, we can exploit this 
symmetry by
 dividing  the jet into ``slices'' across the $z$-axis and
calculating the synchrotron emission and absorption at each slice separately.
Adding the contribution of all slices, we get the total emitted power.

For a relativistic power law distribution of electrons, the emitted synchrotron
power per unit frequency $dP(\nu)$ in a slice in the range ($z$, $z+dz$) is 
(see, e.g., Eq. (6.36) of Rybicki \& Lightman 1979)
\be
dP(\nu)=C_{\alpha}B^{\frac{\alpha+1}{2}}\gamma_{\rm min}^{\alpha-1}n
\sqrt{1-v_{\parallel}^2/c^2}\nu^{\frac{1-\alpha}{2}}R^2dz,
\label{pslice}
\ee
where $C_{\alpha}$ depends only on the index $\alpha$ (defined in Eq. (6)), while
$B=B(z)$, $n=n(z)$, and $R=R(z)$ are given by Eqs. (\ref{bfield}), (\ref{density}), and
(\ref{Rz}), respectively. 

The synchrotron radiation is strongly self-absorbed below a characteristic 
frequency (the turn-over frequency) $\nu_{\rm t}$. Synchrotron 
absorption coefficient is given by the expression (see Eq. (6.53) in Rybicki \&
Lightman 1979)
\be
a_{\nu}=A_{\alpha}n\gamma_{\rm min}^{\alpha-1}B^{\frac{\alpha+2}{2}}\nu^{-\frac{\alpha+4}
{2}},
\label{a_nu}
\ee
where $A_{\alpha}$ depends only on the index $\alpha$.
The turn-over frequency in a slice of the jet can be estimated as the frequency for which the
optical depth to synchrotron absorption, across a radius of this slice
becomes unity, i.e. $a_{\nu_{\rm t}}R(z)\simeq 1$. Solving this
expression for $\nu_{\rm t}$ and using Eqs. (\ref{density}), (\ref{Rz}),
(\ref{bfield}), and (\ref{a_nu}) we have
\be
\nu_{\rm t}=A_{\alpha}^{-\frac{2}{\alpha+4}}n_0^{-\frac{2}{\alpha+4}}
\gamma_{\rm min}^{\frac{2\alpha-2}{\alpha+4}}B_0^{\frac{\alpha+2}{\alpha+4}}
\big(\frac{z_0}{z}\big)^{\frac{\alpha+3}{\alpha+4}}.
\label{nu_t}
\ee   

In estimating the turn-over frequency by setting
$a_{\nu_{\rm t}}R(z)\simeq 1$, we actually assume that the jet thickness $R(z)$
is much less than its height $z$. This does not hold close to the base
of the jet where $z<R(z)$. From Eq. (2) we can verify that $z\gg R(z)$ when
$z\gg 10^3 r_g$. Since the jet extends up  $H=5\cdot 10^8 r_g$, our assumption holds
along most of the jet. We will return to this discussion at the end of Sect. 5.2.     

Before proceeding to a more detailed calculation of the emitted spectrum,
we will calculate the slope of the spectrum over a large range of frequencies.
It is clear from Eq. (\ref{nu_t}) how the turn-over frequency drops with
the height $z$ of the slice of the jet. Since $z$  extends over many orders
of magnitude from $z_0$ up to $H$, there is a large range of frequencies $\nu$
for which part of the jet is optically thin and the rest optically thick.
The height $z_{\nu}$ of the transition from the optically thin region to the optically
thick one depends on the frequency and is given by Eq. (\ref{nu_t})
if we solve for $z$ and call it $z_{\nu}$, namely  
\be
z_{\nu}\propto \nu^{-\frac{\alpha+4}{\alpha+3}}.
\label{z_nu}
\ee

If, for the moment, we take into account only the optically thin part of 
synchrotron emission, to calculate the power emitted at frequency $\nu$,
one has to integrate Eq. (\ref{pslice}) from $z_\nu$ to $H$ and use Eq. (\ref{z_nu}) to
arrive to the scaling
\be
P(\nu)\propto \nu^{\frac{\alpha-1}{2\alpha+6}}.
\label{radios}
\ee 

The slope $\frac{\alpha-1}{2\alpha+6}$ depends rather weakly on $\alpha$ and  
for $3\simless \alpha \simless 5$, the slope is $\sim 0.2$. This result, which will
be verified by the more detailed calculation of Sect. 5.2,
shows that the emitted spectra of our model agree very well with the 
flat to inverted radio spectra that are typically observed
from the steady jets of black hole candidates in the low/hard state.

\begin{figure}
\resizebox{\hsize}{!}{\includegraphics[angle=270]{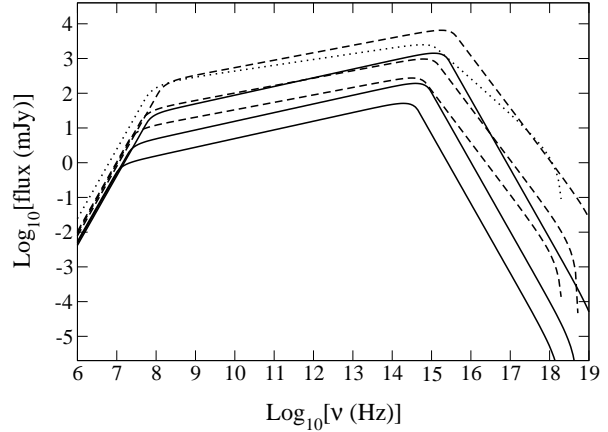}}
\caption[]{The emitted  synchrotron flux (in mJy) as observed at a distance
of 1 kpc from our jet is presented for different values of the index $\alpha$ and
the magnetic field strength at the base of the jet $B_0$. The rest of the parameters
are kept at their reference values (see Sect. 3). The dotted curve corresponds 
to $\alpha=3$ and $B_0=2\cdot 10^5$ G. The dashed curves correspond, in order
of increasing flux, to $\alpha=4$ and $B_0=2\cdot 10^5$, $5\cdot 10^5$, and
$2\cdot 10^6$G respectively. The solid curves correspond to $\alpha=5$ and
$B_0=2\cdot 10^5$, $5\cdot 10^5$, and $2\cdot 10^6$G in order of increasing flux.
The spectrum in the radio is almost flat with spectral index that 
depends weakly on $\alpha$ (see Eq. (\ref{radios})). For frequencies above 
$\sim 10^{15}$Hz
the whole jet is optically thin to synchrotron absorption and has 
slope $(1-\alpha)/2$.     
} 
\label{radio}
\end{figure}

\subsection{The radio spectrum}

Knowing the power emitted per unit frequency in each slice of the jet (Eq. 
(\ref{pslice})) and the turn-over frequency as a function of $z$ (Eq. (\ref{nu_t})), 
the spectrum of the radiation emitted by this slice can be approximated by the
optically thin emission for $\nu>\nu_{\rm t}$ that is smoothly connected to
the optically thick one for $\nu<\nu_{\rm t}$. Adding the contribution
of each slice of the jet from $z_0$ to $H$, we have the total synchrotron 
emission in the rest frame of the jet, $P(\nu)$. 

For two symmetric jets streaming in opposite directions, the total flux that reaches the observer 
located at distance $d$ from the black hole in a direction that makes  
angle $\theta$ with the axis of the 
approaching jet is (see, e.g., Mirabel \& Rodr\'{\i}guez 1999)
\be
f(\nu)=\frac{P(\nu)}{4\pi d^2}(\delta_{\rm apr}^{2-s}+\delta_{\rm rec}^{2-s}),
\label{flux}
\ee
where $s$ is the energy spectral index $\Delta \log [P(\nu)]/\Delta \log (\nu)$.
The Doppler formulae for the approaching, $\delta_{\rm apr}$, and the receding,
$\delta_{\rm rec}$, jets are given by
\be
\delta_{\rm apr}=\frac{1}{\gamma_{\parallel}(1-v_{\parallel}/\cos \theta)},
\label{apr}
\ee
\be             
\delta_{\rm rec}=\frac{1}{\gamma_{\parallel}(1+v_{\parallel}/\cos \theta)}.
\label{rec}
\ee
Finally, $\gamma_{\parallel}=1/\sqrt{1-v_{\parallel}^2/c^2}$ is the bulk 
Lorentz factor of the jet.

For $\cos \theta \simeq 0.5$ and $v_{\parallel}=0.8 c$,
the Doppler shifts are $\delta_{\rm apr}\simeq 1$, $\delta_{\rm rec}\simeq 0.4$.
So, for an intermediate range of viewing angles of our mildly relativistic 
jet, the contribution of the receding jet to the total emitted 
synchrotron flux is rather small (see Eq. (\ref{flux})).
In principle, we should have distinguished between the photon frequency
in the observer and in the comoving frame. The two frequencies scale with
a Doppler factor $\delta_{\rm apr}$. Since, $\delta_{\rm apr} \simeq 1$ in our
study, this distinction is not crucial and has been avoided. 

In Fig. \ref{radio} we plot the synchrotron flux that reaches the
observer, for a source that is located at distance $d=1$ kpc, for
different values of $\alpha$ and $B_0$. The rest of the parameters are
kept at their reference values. Several comments are in order
in view of this plot:

i) The low-frequency part of the spectrum ($10^8-10^{15}$Hz) has a power-law 
form with slope given by Eq. (\ref{radios}). 

ii) The low-frequency cutof\mbox{}f (at $\sim 10^8$ Hz) is given by the value
of the turn-over frequency at the outer edge of the jet at height $H$, while
the high frequency cutof\mbox{}f (at $\sim 10^{15}$Hz) is 
close to the turn-over frequency at the base of the jet $z_0$. This 
break typically lies close to the optical region.

iii) For frequencies $\simmore 10^{15}$ Hz, the whole volume of the
jet is optically thin to synchrotron absorption and the spectrum
has a power law shape with slope $-(\alpha-1)/2$ (see Eq. [\ref{pslice}]).
The high frequency spectrum steepens further as a result of
the fact that the energy distribution of the electrons
extends up to $\gamma_{\rm max}$. Equation (\ref{pslice}) is accurate
up to frequency $\nu=(3/2)\gamma_{\rm max}^2\nu_{\rm L}$,
where $\nu_{\rm L}=eB/(2\pi m_ec)$ is the Larmor frequency.

iv) When normalized to a distance of 1 kpc, the radio emission
of black hole binaries in the low/hard state reaches fluxes of up       
to $\sim$  a few 100 mJy (see Fig. 6 in Gallo et al. 2003). Magnetic
field strengths of the order of $10^6$ G are required at the
base of the jet to explain the most luminous (in the radio) sources
for $3\simless \alpha \simless 5$.   
 
The part of the spectrum close to the high-frequency
break of the flat part (close to $10^{15}$ Hz) is rather approximate. 
Close to the base of the jet the characteristic radius of the 
jet is comparable or even larger than its height, making our
analysis inaccurate for this region.
Furthermore, close to the base of the jet, the electron scattering
optical depth is high enough that it cannot be neglected.
In fact, the synchrotron flux emitted close to the base of the jet may
contribute significantly to the soft photon input that is 
Comptonized by the jet. In the next Section where comparison between the spectra 
calculated from the model and the observational data is made, the
synchrotron photons are also taken into account as a soft photon source
to be Comptonized in the jet.

\section{Applying the model to XTE J1118+480 and Cygnus X-1}  
           
The synchrotron flux, as predicted by our jet model, peaks at $\sim 10^{15}$
Hz. Most of the contribution at this frequency comes from the
lower region of the jet. On the other hand, synchrotron photons that are emitted
in this region have a significant probability of being scattered by
 electrons once or more times. So, synchrotron emission must also be taken into account
along with the disk emission as a soft photon source.

For the black-hole binary XTE J1118+480, simultaneous (or nearly simultaneous) 
observations have been conducted on multiple occasions at radio, infrared, optical, 
UV, EUV, and X-ray wavelengths (Hynes et al. 2000; McClintock et al. 2001b; Frontera 
et al. 2001), making it an ideal source to test the model in the whole electromagnetic 
spectrum. In Fig. 4 the most complete spectral energy distribution of XTE J1118+480
is shown (associated with the so-called ``epoch 2''), where the radio data are
from Fender et al. (2001; we do not include the observational point at 350 GHz measurement
 which was not done simultaneously with the others) and the infrared to X-ray data from 
McClintock et al. (2001a). Since the EUV spectrum depends sensitively on the assumed 
$N_{\rm H}$ which is not well constrained but probably lies in the range 
$1.0-1.3\times 10^{20} \rm cm^{-2}$ (McClintock et al. 2001b), we choose to plot the data
corrected for $N_{\rm H}=1.0\times 10^{20} \rm cm^{-2}$ and $N_{\rm H}=1.3\times 10^{20} 
\rm cm^{-2}$. The spectrum of XTE J1118+480 has been fitted by an accretion-jet model
(Yuan et al. 2005; see also Esin et al. 2001), a synchrotron model (Markoff et al. 2001)
and a thermal-Comptonization model (Frontera et al. 2001); for a comparison among different
models see Chaty et al. (2003). Notice that in the ADAF and corona models the presence of
a jet is needed to explain the observed emission in the infrared and radio wavelengths.

In Fig. 4, the spectrum from our model is also shown. 
The parameters that have been used are  $R_0=70 r_g$, $\alpha=3.5$, $T_{bb}=10$ eV 
and $B_0=10^5$ G, while the rest of the parameters have been kept to their reference 
values. The overall agreement between the calculated and observed spectra is very good. 
Both disk and synchrotron photons from the jet 
contribute an important fraction of the soft flux to be Comptonized in the jet. 

 It is also clear from Fig. 4 that in this model the optically thin synchrotron 
emission has only a minor contribution to the X-ray spectrum in contrast to the jet model 
of Markoff et al. (2001). The main reason for this discrepancy is the steeper distribution 
function of the emitting electron that we favor (which results in steeper optically thin 
synchrotron spectra) and the higher electron densities in our jet model (that result in 
efficient inverse Compton scattering in the jet). Recently,
Heinz (2004)  has shown that, adding radiative
cooling to the scale invariance formalism of Heinz \& Sunyaev (2003), which is in 
accordance with the ``fundamental plane'' of black hole activity (Merloni et al. 2003),
 Compton scattering is favored over synchrotron as the emission mechanism
of the hard X-rays for the majority of these sources. 

Besides its broad-band spectral information, XTE J1118+480 has also been extensively
observed in the temporal domain where a large number of interesting properties has been revealed 
for the lightcurves in different energy bands. For example, quasi-periodic oscillations have 
been observed with the same frequency in the optical, UV and X-rays (Hynes et al. 2003), 
and the correlation between emission at different wavelengths
 is clear and puzzling (Kanbach et al. 2001; Spruit \& Kanbach 2002; Hynes et al. 2003).
It is interesting (though not within the scope of this work) to see whether our jet model 
is able to reproduce these observations.

\begin{figure}
\resizebox{\hsize}{!}{\includegraphics[angle=270]{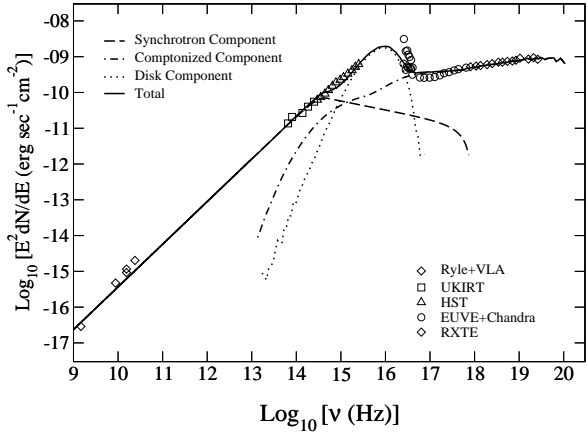}}
 \caption[]{ The broad-band spectrum from our model for $R_0=70 r_g$, $\alpha=3.5$, 
$T_{bb}=10$ eV, and $B_0=10^5$ G  
with the rest of the parameters kept at their reference values (solid curve) 
is overplotted with the radio to X-ray spectrum of XTE J1118+480. The dashed, dash-dotted 
and dotted curves show the synchrotron, 
Comptonized and disk components respectively. Notice that the EUV spectrum depends sensitively
on the assumed $N_{\rm H}$ and is derived for $N_{\rm H}=1.0\times 10^{20} \rm cm^{-2}$ and
$N_{\rm H}=1.3\times 10^{20} \rm cm^{-2}$ (McClintock et al. 2001b).} 
\label{j1118}
\end{figure}

\begin{figure}
\resizebox{\hsize}{!}{\includegraphics[angle=270]{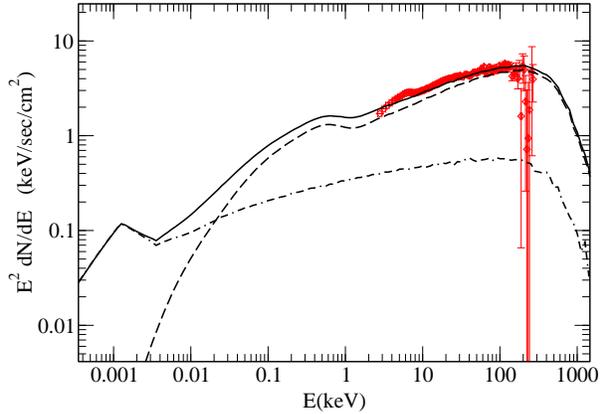}}
 \caption[]{The photon-number spectrum multiplied by $E^2$ 
for the reference values of  our jet model (solid curve) is overplotted with the 
spectrum of Cygnus X-1 as observed by RXTE in December 1996. The dashed and dashed-dotted 
curves show the disk+Comptonized disk and synchrotron self Compton components respectively.}

\label{spectrum1}
\end{figure}

 We now proceed to apply the model to the December 1996 RXTE observations
of Cygnus X-1 when the source was in the low/hard state. For the reference values 
of the parameters, the simulated spectra are 
in rather good agreement with the observed ones (see Fig. \ref{spectrum1}). On the 
other hand, the model underpredicts the emitted flux for $E\simless 30$ keV.

This is the region where the reflection component, considered to come from hard x-rays
that are reprocessed or reflected by the thin disk, is expected to contribute most.
In this work, the reflection component has not been included in the calculation. 
Nevertheless, we have computed (using the Monte Carlo simulation for the reference
values of the parameters) that, assuming that the disk is infinitely 
thin and that the jet is at an angle of 90 degrees to the disk plane, $\sim 4$\% 
of the X-ray flux hits the disk. If the disk is flared or warped (e.g. Dubus
et al. 1999) then $\sim 8$\% of the X-ray flux reaches the disk (taking an
``effective'' $h/R=0.2$, where $h$ is the disk half-thickness and r the radial 
distance). The fraction of X-rays that hits the disk can increase
further if the jet is misaligned with respect to the outer disk (Maccarone 2002). 
Thus the reflection component can be rather strong and depends sensitively on the
accretion-jet geometry and the velocity of the jet. More detailed modeling is needed to
accurately compute the strength of this component as predicted by the model.

\section{Conclusion}

The hard X-ray emission in black-hole binaries is usually modeled to 
come from inverse Compton scattering of soft photons in a hot corona (e.g.
Poutanen \& Fabian 1999) that 
lies above and below a thin disk (Shakura \& Sunyaev 1973) or in a 
two-temperature flow (Shapiro et al. 1976). Here we explore an alternative 
picture where the inverse Compton scattering takes place in the jet whose 
presence, whenever the source is in the low/hard state, is well established 
(Fender 2001; Stirling et al. 2001; Gallo et al. 2003).

In Paper I, it was shown that, if the density profile of the jet
drops inversely proportionaly to distance,  power-law spectra
with photon number index 1.5 - 2, a cutof\mbox{}f at a few 100 keV and
phase lags almost independent of Fourier frequency can be derived
for a wide range of the model parameters. Furthermore in Paper II,
assuming that the more energetic electrons lie at small polar distances in
the jet, both the hardening of the high-frequency power spectrum
and the narrowing of the autocorrelation function with increasing photon energy 
could be reproduced, in agreement with observations of Cygnus X-1.

Here, we modified this jet model to include a steep power-law electron
distribution (with $3\simless \alpha\simless 5$,  that is a result 
of fast inverse Compton -and synchrotron- cooling of the electrons) in the jet 
and checked that all the results derived in the X-ray region still hold. 
We also modeled the emission that comes from the jet at 
longer wavelengths in terms of synchrotron emission of the same population of 
electrons and found that the flat to inverted radio spectra, observed in black-hole
candidates in the low/hard state, are a natural outcome of the model.
Magnetic fields of the order of $10^5-10^6$G at the base of the
jet are needed for the emitted power in the radio to match the typically observed one.         

The flat radio spectra have been shown to extend up to about the optical
wavelengths, with the base of the jet  emitting mostly in this
energy band. This synchrotron flux may be strong enough to contribute
significantly to the soft photon input that is Comptonized in the jet.
 The model has been applied to the broad band spectral energy distribution
of J1118+480 and Cygnus X-1 showing good agreement with observations.          
                      
{\it Acknowledgments.} I would like to thank Nick Kylafis for very useful
discussions and comments on this manuscript.  I also wish to thank the anonymous
referee for comments that greatly improved and clarified the manuscript.  This research has been 
supported by the Program ``Heraklitos'' of the Ministry of Education of Greece.


\begin{thebibliography}{}

\bibitem{} Abramowicz, M. A., Chen, X. M., Kato, S., Lasota, J. P., \& Regev,
O. 1995, ApJ, 438, L37
\bibitem{} Band, D., \& Grindlay, J. E. 1986, ApJ, 311, 595
\bibitem{} Blandford, R. D., \& K\"onigl, A. 1979, ApJ, 232, 34
\bibitem{} Chaty, S., Haswell, C. A., Malzac, J., et al. 2003, MNRAS, 346, 689   
\bibitem{} Corbel, S., Fender, R. P., Tzioumis, A. K., et al. 2000, A\&A, 359, 251   
\bibitem{} Corbel, S. \& Fender, R. P. 2002, ApJ, 573, 35
\bibitem{} Corbel, S., Nowak, M. A., Fender, R. P., Tzioumis, A. K., \& Markoff,
S. 2003, A\&A, 400, 1007
\bibitem{} Dhawan, V., Mirabel, I. F., Rodr\'\i guez, L. F. 2000, ApJ, 543, 373
\bibitem{} Dubus, G., Lasota, J., Hameury, J., \& Charles, P. 1999, MNRAS, 303, 139
\bibitem{} Esin, A. A., McClintock, J. E., \& Narayan, R. 1998, ApJ, 500, 523
\bibitem{} Esin, A. A., McClintock, J. E., Drake, J. J., et al. 2001, ApJ, 555, 483
\bibitem{} Falcke, H., K\"{o}rding, E., \& Markoff, S. 2004, A\&A, 414, 895 
\bibitem{} Fender, R. P. 2001, MNRAS, 322, 31
\bibitem{} Fender, R. P. 2003, in {\em Compact Stellar X-Ray Sources}, eds. W. H. D.
Lewin, \& van der Klis, Cambridge University Press
\bibitem{} Fender, R. P., Pooley, G. G., Durouchoux, P., Tilanus R. P. J., \&
Brocksopp, C. 2000, 312, 853
\bibitem{} Fender, R. P., Hjellming, R. M., Tilanus, R. P. J., et al. 2001, MNRAS, 322 , L23
\bibitem{} Ford, E. C., van der Klis, M., M\'endez, M., van Paradijs, J.,
\& Kaaret, P. 1999, ApJ, 512, L31 
\bibitem{} Frontera, F., Zdziarski, A. A., Amati, L., et al. 2001, ApJ, 561, 1006
\bibitem{} Fuchs, Y., Mirabel, I. F., Rodr\'\i guez, L. F., et al. 2003, A\&A, 409, L35 
\bibitem{} Galeev, A. A., Rosner, R., \& Vaiana, G. S. 1979, ApJ, 229, 318
\bibitem{} Gallo, E., Fender, R. P., \& Pooley, G. G. 2003, MNRAS, 344, 60 
\bibitem{} Georganopoulos, M., Aharonian, F. A., \& Kirk, J. G. 2002, A\&A,
388, L25
\bibitem{} Giannios, D., Kylafis, N. D., \& Psaltis, D. 2004, A\&A, 425, 163 (Paper II)
\bibitem{} Haardt, F., \& Maraschi L. 1993, ApJ, 413, 507
\bibitem{} Haardt, F., Maraschi L., \& Ghisellini, G. 1994, ApJ, 432, L95
\bibitem{} Hannikainen, D. C., Hunstead, R. W., Campbell-Wilson, D., \& Sood,
R. K. 1998, A\&A, 337, 460
\bibitem{} Heinz, S. 2004, MNRAS, 355, 835
\bibitem{} Heinz, S., \& Sunyaev, R. A. 2003, MNRAS, 343, 59
\bibitem{} Hjellming, R.M., \& Han, X. 1995, {\em Radio properties of X-ray binaries.}
In : Lewin, W. H. G., van Paradijs, J., van der Heuvel, E. P. J., (Eds.), X-ray binaries,
Cambridge University Press, Cambridge, p. 308 
\bibitem{} Hynes R. I., Mauche, C., Shrader, C., Cui, W., \& Chaty, S. 2000, ApJ, 539, L37
\bibitem{} Hynes R. I., Haswell, C. A., Cui, W., et al. 2003, MNRAS, 345, 292 
\bibitem{} Ichimaru, S. 1977, ApJ, 214, 840
\bibitem{} Kanbach, G., Straubmeier, C., Spruit, H. C. \& Belloni, T. 2001, Nature, 414, 180
\bibitem{} Kylafis, N. D., Giannios, D., \& Psaltis, D. 2004, in {\em X-ray Timing
2003: Rossi and Beyond}, eds. P. Kaaret, F.K. Lamb, \& J.H. Swank, American Institute of
Physics, Vol. 714, p. 101 
\bibitem{} Levinson, A., \& Blandford, R. 1996, ApJ, 456, L29
\bibitem{} Maccarone, T. J. 2002, MNRAS, 336, 1371
\bibitem{} Maccarone, T. J., Coppi, P. S., \& Poutanen, J. 2000, ApJ, 537, L107
\bibitem{} Markoff, S., Falcke, H., \& Fender, R. 2001, A\&A, 372, L25
\bibitem{} Markoff, S., Nowak, M. A., Corbel, S., Fender, R., \& Falcke, H. 2003,
A\&A, 397, 645
\bibitem{} Markoff, S., \& Nowak, M. A. 2004, ApJ, 609, 972
\bibitem{} Mart\'{\i}, J., Mirabel, I. F., Rodr\'\i guez, L. F., \& Smith, I. A. 2002, 386, 571 
\bibitem{} Merloni, A., Heinz, S., \& di Matteo, T. 2003, MNRAS, 345, 1047 
\bibitem{} McClintock, J. E., \& Remillard, R. E. 2005, in {\em Compact Stellar X-ray Sources},
eds. W. H. G. Lewin, M. \& M. van der Klis, Cambridge University Press, Cambridge, 
astro-ph/0306213   
\bibitem{} McClintock, J. E., Garcia, M. R., Caldwell N., Falco, E. E., Garnavich, P. M.,
\& Zhao, P. 2001a, ApJ, 551, L147
\bibitem{} McClintock, J. E., Haswell, C. A., Garcia, M. R., et al. 2001b, ApJ, 555, 477
\bibitem{} Merloni, A., Heinz, S., \& di Matteo, T. 2003, MNRAS, 345, 1047   
\bibitem{} Mirabel, I. F., \& Rodr\'\i guez, L. F. 1999, ARA\&A, 37, 409
\bibitem{} Mirabel, I. F., Rodr\'\i guez, L. F., Corbier, B., Paul, J., \& Lebrun, F.
1992, Nature, 358, 215 
\bibitem{} Mirabel, I. F., Dhawan, V., Chaty S., et al. 1998, A\&A, 330, L9
\bibitem{} Mitsuda, K., Inoue, H., Koyama, K., et al. 1984, PASJ, 36, 741
\bibitem{} Miyamoto, S., Kitamoto, S., Mitsuda, K., \& Dotani, T. 1988,
Nature, 336, 450
\bibitem{} Narayan, R., \& Yi, I. 1994, ApJ, 428, L13
\bibitem{} Nowak, M. A. 1995, PASP, 107, 1207
\bibitem{} Nowak, M. A., Vaughan, B. A., Wilms, J., Dove, J. B., \& Begelman,
M. C. 1999, ApJ, 510, 874  
\bibitem{} Poutanen, J. 1998, in {\it Theory of Black Hole Accretion Disks}, Cambridge
Univ. Press, Cambridge, p. 100 
\bibitem{} Poutanen, J., \& Fabian, A. C. 1999, MNRAS, 306, L31 
\bibitem{} Pozdnyakov L. A., Sobol I. M., \& Sunyaev R. A. 1983,
Astrophys. \& Space Phys. Rev. 2, 189
\bibitem{} Rees, M. J., Phinney, E. S., Begelman, M. C., \& Blandford, R. P.
1982, Nature, 295, 17 
\bibitem{} Reig, P., Kylafis, N. D., \& Giannios, D. 2003, A\&A, 403, L15 (Paper I)
\bibitem{} Revnivtsev, M., Gilfanov, M., \& Churazov, E. 2000, A\&A, 363, 1013
\bibitem{} Romero, G. E., Kaufman Bernad\'o, M. M., \& Mirabel, F. 2002,
A\&A, 393, L61
\bibitem{} Rybicki G. B., \& Lightman A.P. 1979, {\em Radiative Processes in Astrophysics},
Wiley, New York
\bibitem{} Stern, B. E., Poutanen, J., Svensson, R., Sikora, M., \& Begelman,
M. C. 1995, ApJ, 449, L13
\bibitem{} Stirling, A. M., Spencer, R. E., de la Force C.J., et al. 2001, MNRAS,
327, 1273
\bibitem{} Shakura, N. I., \& Sunyaev, R. A. 1973, A\&A, 24, 337
\bibitem{} Shapiro, S. L., Lightman, A. P., \& Eardley, D. M. 1976, ApJ, 204, 187
\bibitem{} Spruit, H. C., \& Kanbach, G. 2002, A\&A, 391, 225 
\bibitem{} Sunyaev, R. A., \& Titarchuk, L. G., 1980, A\&A, 86, 121
\bibitem{} Tanaka, Y. \& Lewin, W. H. G. 1995,  in {\em X-ray Binaries}, eds. W. H. G. Lewin 
J. van Paradijs, \& E. P. j. van den Heuvel, Cambridge University Press, Cambridge, p. 126-174
\bibitem{} Vadawale, S. V., Rao, A. R., \& Chakrabarti, S. K. 2001, A\&A, 372, 793
\bibitem{} van der Klis, M. 1995, in {\em The lives of neutron stars},
Kluwer Academic Publishers., Eds. M. A. Alpar, \"U. Kiziloglu, \& J. van Paradijs,
NATO ASI Series C450.
\bibitem{} Yuan, F., Cui, W., \& Narayan, R. 2005, ApJ, in press, astro-ph/0407612 

\end{thebibliography}
\end{document}